\documentclass[12pt,preprint]{aastex}

\lefthead{Skinner et al.}
\righthead{WR 110}

\begin{document}

\title{{\em XMM-Newton} Detection of Hard X-ray Emission \\ 
            in the Nitrogen-type Wolf-Rayet Star WR 110  }

\author{Stephen L. Skinner}
\affil{CASA, Univ. of Colorado, Boulder, CO 80309-0389 }

\author{Svetozar A. Zhekov \altaffilmark{1}}
\affil{JILA, Univ. of Colorado, Boulder, CO 80309-0440 }
\altaffiltext{1}{On leave from Space Research Institute,
Sofia, Bulgaria}

\author{Manuel G\"{u}del}
\affil{Paul Scherrer Institute, W\"{u}renlingen and Villigen,
CH-5232 Switzerland}

\and 

\author{Werner Schmutz}
\affil{Physikalisch-Meteorologisches Observatorium Davos,
Dorfstrasse 33, 
CH-7260 Davos Dorf, Switzerland}

% Notice that each of these authors has alternate affiliations, which
% are identified by the \altaffilmark after each name.  The actual alternate
% affiliation information is typeset in footnotes at the bottom of the
% first page, and the text itself is specified in \altaffiltext commands.
% There is a separate \altaffiltext for each alternate affiliation
% indicated above.

% The abstract environment prints out the receipt and acceptance dates
% if they are relevant for the journal style.  For the aasms style, they
% will print out as horizontal rules for the editorial staff to type
% on, so long as the author does not include \received and \accepted
% commands.  This should not be done, since \received and \accepted dates
% are not known to the author.
%
% Define symbol \ltsimeq
\newcommand{\ltsimeq}{\raisebox{-0.6ex}{$\,\stackrel{\raisebox{-.2ex}%
{$\textstyle<$}}{\sim}\,$}}
\begin{abstract}
We have used the excellent sensitivity of XMM-Newton to obtain the
first high-quality X-ray spectrum of a Wolf-Rayet (WR) star which 
is not known to be a member of a binary system. Our target, the  
nitrogen-type star WR 110 (= HD 165688) was also observed and 
detected with the VLA at four  different frequencies. The radio
flux density increases with frequency according to a power law
S$_{\nu}$ $\propto$ $\nu^{+0.64 \pm 0.10}$, in very good agreement
with the behavior expected for free-free wind emission. The radio data
give an ionized mass-loss rate  $\mathrm{\dot{M}}$ =
4.9 $\times$ 10$^{-5}$ M$_{\odot}$ yr$^{-1}$ for an
assumed spherical constant-velocity wind. 

The undispersed CCD X-ray spectra reveal strong emission
lines from He-like ions of Mg, Si and S. The emission
measure distribution shows a dominant contribution from cool plasma
with a characteristic temperature kT$_{cool}$ $\approx$ 0.5 keV
($\approx$6 MK).
Little or no excess absorption of this cool component above the
value expected from the visual extinction is present. We conclude
that the bulk of the cool plasma detected by XMM-Newton lies at hundreds 
of stellar radii or more if the wind is  approximately spherical and 
homogeneous, but it could lie closer to the star if the wind is clumped.
If the cool plasma is due to instability-driven wind shocks
then  typical shock velocities are
v$_{s}$ $\approx$ 340 - 550 km s$^{-1}$ and the  average filling
factor of X-ray emitting gas in the wind is no larger than 
$f$ $\sim$ 10$^{-6}$.

A surprising result is the unambiguous detection of a hard X-ray component
which is clearly seen in the hard-band images and the spectra.  
This hard
component accounts for about half of the observed flux and can be
acceptably fitted by a hot optically thin thermal plasma or a power-law
model. If the emission is thermal, then a temperature
kT$_{hot}$ $\geq$ 3 keV is derived. Such high temperatures are not
predicted by current  instability-driven wind shock models and a
different mechanism is thus required to explain the hard X-rays.
We examine several possible mechanisms and show that the hard 
emission could be accounted for by the WR wind shocking onto a close
stellar companion which has so far escaped detection. However, until
persuasive evidence for binarity is found we are left with the intriguing 
possibility that the hard X-ray emission is produced entirely by the
Wolf-Rayet star.
\end{abstract}

% The different journals have different requirements for keywords.  The
% keywords.apj file, found on aas.org in the pubs/aastex-misc directory, 
% contains a list of keywords used with the ApJ and Letters.  These are 
% usually assigned by the editor, but authors may include them in their 
% manuscripts if they wish. 


% That's it for the front matter.  On to the main body of the paper.
% We'll only put in tutorial remarks at the beginning of each section
% so you can see entire sections together.

% In the first two sections, you should notice the use of the LaTeX \cite
% command to identify citations.  The citations are tied to the
% reference list via symbolic KEYs.  We have chosen the first three
% characters of the first author's name plus the last two numeral of the
% year of publication.  The corresponding reference has a \bibitem
% command in the reference list below.
%
% Please see the AASTeX manual for a more complete discussion on how to make
% \cite-\bibitem work for you.   
\newpage

\section{Introduction}
The {\em XMM-Newton} and {\em Chandra} observatories are now providing
dramatic new  high-resolution X-ray images and spectra of Wolf-Rayet
(WR) stars. The improvements in sensitivity and spectral resolution 
offered by these observatories have effectively opened a new window
for the study of WR winds and atmospheres. High-resolution spectra
can potentially provide information on temperature structure, 
electron densities, chemical composition, and velocity profiles.
A fundamental question that can now be addressed is the origin of the X-ray
emission in WR stars, which has been a long-standing puzzle.

The most detailed X-ray studies of WR stars to date have focused on
X-ray bright WR $+$ OB binaries, for which high signal-to-noise
spectra can be obtained in relatively short exposures. For example,
high resolution {\em Chandra} and {\em XMM-Newton} grating spectra of 
$\gamma^2$ Velorum (WC8 $+$ O7.5) have recently been obtained
(Skinner et al. 2001; Schmutz et al. 2002). In contrast,
single WR stars without known companions are typically much fainter
in X-rays and even moderate quality CCD spectra have been difficult
to acquire. The capability to obtain CCD spectra of such fainter
objects  has now become available
with  {\em XMM-Newton}, which  provides the largest effective area ever
achieved in an imaging X-ray telescope (Jansen et al. 2001). 
We present here the results of an {\em XMM-Newton} observation of
the nitrogen-type WN 5-6 star  WR 110 (= HD 165688), providing the
best X-ray spectrum obtained so far of a WR star with no 
known companion.

Several unanswered questions motivate the study of X-ray emission from
single WR stars. At a basic level, we would like to identify the physical
process that is responsible for heating the plasma to X-ray emitting 
temperatures of several million K, and we  would like to know if
the same process is responsible for X-ray emission in both WR and OB
stars. It is commonly assumed that the single WR and OB stars do
emit X-rays by the same process and the emission is usually attributed
to shocks that form in their powerful winds as a result of 
line-driven instabilities  (Lucy \& White 1980; Lucy 1982). 
The emergent X-ray
emission is expected to be soft (kT $\leq$ 0.5 keV) and weakly absorbed.
This mechanism has most generally been applied to OB stars, but
radiative transfer models suggest that line-driven flow instabilities
can also form in  WR winds (Gayley \& Owocki 1995).

Given the traditional belief that single WR and OB stars have a common
X-ray emission mechanism, it is somewhat surprising that statistical
X-ray correlations known to exist in OB stars are apparently not 
present in WR stars.
Wessolowski (1996) analyzed ROSAT data for a sample of 61 putatively
single WN-type stars and found no evidence for the correlation 
between  X-ray luminosity L$_{x}$ and bolometric luminosity 
L$_{bol}$ that is well-documented in single O-type stars (Bergh\"{o}fer
et al. 1997). Furthermore, no correlation between L$_{x}$ and wind momentum
loss rate $\mathrm{\dot{M}}$v$_{\infty}$ was seen in the WN star sample, 
in contrast to
what might be expected if the X-rays arise in shocked winds.
It is not yet known if these results reflect true differences between
WR and OB star X-ray  processes, or whether other factors such
as undetected binarity might be responsible.

X-ray spectra of single WR stars are needed to address the basic
question of the origin of their X-ray emission, and also to refine X-ray
emission models of more complex  WR $+$ OB binaries. The X-ray
emission of such binaries may be the superposition of intrinsic
emission from the stars themselves plus an extrastellar component
arising from colliding wind shocks between the two stars
(Prilutskii \& Usov 1976; Usov 1992). With the possible
exception of WR 147 (Pittard et al. 2001),  X-ray telescopes 
lack sufficient angular resolution to spatially resolve the stellar
and colliding wind components. Thus, previous studies have assumed
that the hot X-ray emission (kT $\geq$ 1 keV) detected in 
WR $+$ OB spectroscopic binaries is due solely to  colliding wind 
shocks, and that any intrinsic stellar contribution is negligible.
Obviously this assumption needs to be tested by acquiring X-ray
spectra of single WR and OB stars since a hot intrinsic stellar
component (if present) could masquerade as colliding wind shock
emission in spectroscopic binaries. 

WR 110 is an ideal target for exploring the X-ray properties of
single WR stars since it lies in an uncrowded field and there
is no present evidence for binarity (Table 18 of van der Hucht 2001, 
hereafter  vdH01). 
It also had the highest X-ray count rate of any single
WR star detected by  {\em ROSAT} (0.016 c s$^{-1}$ in a
8.3 ksec pointed observation, RP200717). Its visual 
magnitude  is V $=$ 9.4 with  moderate extinction A$_{V}$ = 4.08 mag
and an estimated distance of 1.28 kpc (vdH01). The new XMM-Newton results
presented here show that the X-ray emission of WR 110 is dominated
by relatively cool plasma at  kT$_{cool}$ $\approx$ 0.55 keV ($\approx$6 MK),
but a  hotter component is also present. We use the X-ray
properties to place constraints on wind shock models and discuss
possible origins of the  hot component. In addition,
we present the first multifrequency VLA observations of WR 110 which
show that its radio properties are in good agreement with that expected
for free-free emission from an ionized wind. \\

\section{Observations}

\subsection{XMM-Newton Observations }

{\em XMM-Newton}  observed WR110 during a
seven hour interval from 0147 - 0848 UT on 2001 March 22,
yielding $\approx$25 ksec of usable exposure time (Table 1).
The observatory consists of three X-ray telescopes as
described by Jansen et al. (2001). 
Data were acquired with the European Photon Imaging Camera (EPIC)
which provides simultaneous CCD imaging spectroscopy from 
the EPIC PN camera (Str\"{u}der et al. 2001) and two identical 
MOS cameras (MOS-1 and MOS-2; Turner et al. 2001). We utilized
full-window mode and the medium optical blocking filter.
Grating spectrometer data were also obtained but lacked sufficient 
counts for spectral analysis. The PN and MOS cameras provide
a $\approx$30$'$ diameter field-of-view and energy coverage
from $\approx$0.2 - 15 keV, moderate energy
resolution (E/$\Delta$E $\approx$ 20 - 50), and 
$\approx$12$''$ FWHM angular resolution. 

Data reduction followed standard procedures using the 
{\em XMM-Newton} Science Analysis System software (SAS vers. 5.1).
The pipeline processing tasks EMCHAIN and EPCHAIN were 
executed using the latest calibration files and data
were  filtered with EVSELECT to select good event patterns.
Spectra and light curves were extracted from the filtered events lists
using circular regions of radius  $\approx$40$''$.
Background was extracted from source-free regions of the same size on the 
same CCD as the source. Data from MOS-1 and MOS-2 were combined
to increase the signal-to-noise ratio (S/N), thus reducing 
uncertainties in parameters derived from spectral fits and 
light curve analysis. Spectra were analyzed with XSPEC vers. 11
(Arnaud 1996)  using a variety of models including discrete 
temperature approximations (MEKAL, APEC) and differential emission 
measure (DEM) analysis. For the hard component, nonthermal power-law
models were also examined. All models included an absorption component
based on Morrison \& McCammon (1983) cross sections.

\subsection{VLA Observations}

Radio observations of WR 110 were obtained with the 
%***
NRAO \footnote{The National Radio Astronomy Observatory (NRAO) is 
a facility of the National Science Foundation operated
under cooperative agreement by Associated Universities
Inc.}
%***
VLA in a three hour observation on 1999 Dec 26
with the array in B configuration, as summarized in Table 2.
WR110 was observed and detected at four frequencies:
4.86 GHz (6 cm), 8.44 GHz (3.6 cm),
14.94 GHz (2 cm), and 22.46 GHz (1.3 cm). 
The star was observed with the full array at each frequency 
in scans of $\approx$10 - 12 minutes duration interleaved with
scans of the phase calibrator 1733$-$130. The primary flux 
calibrator 3C48 was  observed at each frequency.

Data were edited and calibrated using the AIPS 
\footnote{Astronomical Image Processing System (AIPS)
is a software package developed by NRAO.}
software package. Maps were produced in both total
intensity (Stokes $I$) and circularly polarized
intensity (Stokes $V$) using the AIPS task IMAGR with 
natural weighting so as to maximize sensitivity.
Both peak and total (integrated) fluxes were measured
in cleaned maps using the AIPS tasks TVSTAT (pixel 
summation within a region defined by the  2$\sigma$ contour)
and IMFIT (Gaussian source model). These two methods gave
very good agreement, with typical differences of only
3\% in the computed total flux.

\section{Results }

\subsection{X-ray Properties of WR 110}

\subsubsection{X-ray Images}

Figure 1 shows the inner region of the EPIC-PN image 
within $\approx$3 arc-min of WR110 in both broad-band
(0.3 - 10 keV) and hard-band (6 - 10 keV) energy filters.
WR 110 is clearly detected in both images.
The X-ray position obtained by averaging the
results of all three EPIC cameras (Table 1) is 
offset by only 0.$''$6 from the optical position (vdH01),
providing high confidence that the X-ray source is 
associated with WR 110.

The broad-band PN image yields 1916 raw counts from
WR 110 in 22547 s, while MOS-1 gives 989 counts
(25210 s) and MOS-2 gives 1034 counts (25207 s).
The hard-band PN count rate is 
2.31 $\pm$ 0.47 c ksec$^{-1}$ (6 - 10 keV) and the
hard-band detection is confirmed in the combined MOS
images. In addition, evidence for a hard component is
seen in the spectra (Sec. 3.1.3). It is apparent from
Figure 1 that the hard photons
are localized at the WR 110 position and are not due to
background. A second hard source lies  52$''$ NE of WR 110,
but a SIMBAD search gives no counterpart within 30$''$ of 
the X-ray position of this second source.

\subsubsection{X-ray Light Curves and Timing Analysis}

Figure 2 shows the broad-band (0.3 - 10 keV) light
curve of WR 110   obtained by combining data 
from the MOS-1 and MOS-2 detectors. The average 
count rate in the summed MOS light curve is
0.084 $\pm$ 0.013 (1$\sigma$) c s$^{-1}$. 
A $\chi^2$ test assuming a constant count-rate source
gives a probability of constant count rate P(const) = 0.93
($\chi^2$/dof = 34.6/48; binsize = 512 s, 49 bins). A smaller
binsize of 256 s also gives P(const) = 0.93 ($\chi^2$/dof = 77.7/97;
98 bins). Thus there is no evidence for significant
variability down to a binsize of 256 seconds.  Various 
soft-band and hard-band light curves were also generated
using larger binsizes of 2048 s to obtain an acceptable
number of counts per bin, and a similar analysis showed
no compelling evidence for variability. 

The above results are consistent with the general behavior
of hot stars, which rarely show X-ray variability on timescales
of a few hours (e.g. Bergh\"{o}fer et al. 1997). However, some
WR $+$ OB binaries which exhibit steady X-ray emission on short 
timescales are known to vary on orbital timescales of months to
years (e.g. WR 11 and WR 140). Since long-term X-ray variability
provides one possible means of detecting binarity, continued 
X-ray monitoring of WR 110 might be worthwhile (see also 
Sec. 4.4).

\subsubsection{X-ray Spectra}

The EPIC-PN spectrum shown in Figure 3 reveals prominent emission
lines from the He-like ions Mg XI, Si XIII, and S XV. 
The spectrum also shows a 
rather narrow feature near 0.80 keV which we classify as a possible
detection of Fe XVII and structure near 3.03 keV that is probably
weak S XIV/XV emission. Using  maximum line-power temperatures
T$_{max}$ as a guide (Mewe et al. 1985), we infer a range of plasma 
temperatures from  $\sim$6 MK (Mg XI) up to at least 
$\sim$ 16 MK (S XV). It is thus obvious that the plasma is
{\em not} isothermal and our spectral analysis  focuses on
multi-temperature optically thin plasma models and DEM models.

Our attempts to derive  abundances from the PN and MOS spectra
show significant differences between the values inferred from  
VMEKAL and VAPEC optically thin plasma models. These differences
occur even for elements with strong line emission such as Si
and S, and we suspect that they are at least partially due to
differences in atomic data used by the models. As such, 
reliable abundances for WR 110 cannot yet be determined
from the CCD spectra. Thus, for purposes of spectral fitting,
we adopted the typical WN abundances given in Table 1 of van der Hucht,
Cassinelli, and Williams (1986, hereafter vdH86). These abundances
reflect the chemical evolution that occurs in WN stars as a result of 
CNO-cycle burning, including hydrogen depletion  and the 
enhancement of helium and nitrogen. Although the adopted
abundances cannot be considered definitive, they do
lie within the range of values
reported in the literature for WN stars (Willis 1996).
For comparison, we also give results for spectral fits using
solar abundances (Anders \& Grevesse 1989).

Discrete-temperature models with two temperature components
(2T MEKAL and 2T APEC) converge rapidly to a cool component
at kT$_{cool}$ $\approx$ 0.55 keV and a hot component with an uncertain
temperature. Similarly, DEM models show a dominant contribution
from plasma at kT $\approx$ 0.5 keV as well as a turnup above
4 keV signalling hotter plasma. A representative DEM profile is
shown in Figure 4, obtained from fits of the   background-subtracted
PN spectrum using the C6PVMKL model. This model uses an iterative
algorithm based on Chebyshev polynomials (Lemen et al. 1989).
As can be seen, the dominant component at kT $\approx$ 0.5 
keV is a robust result and is recovered using canonical WN-star
abundances as well as solar abundances.  If the hot component
is excluded, then the spectral fit is unacceptable above 
$\sim$3 keV as shown in the unfolded spectrum in Figure 5.
These spectral analysis results, along with the clear
detection of hard photons in the EPIC images, 
lead us to conclude that the hard emission is real and
associated with WR 110.

Table 1 summarizes the plasma properties deduced from 
discrete-temperature and DEM models of the X-ray spectra
using a single absorption component. These models give
an equivalent neutral hydrogen column density 
N$_{H}$ = 1.05 ($\pm$0.18) $\times$ 10$^{22}$ 
cm$^{-2}$. This value agrees to within the uncertainties with
the value N$_{H}$ = 0.91 ($\pm$0.13) $\times$ 10$^{22}$ cm$^{-2}$
computed from the average visual extinction A$_{V}$ = 4.08 $\pm$ 0.58
(Gorenstein 1975). This average A$_{V}$ is slighter larger than
that quoted in Table 24 of vdh01 since we have used a revised value 
A$_{V}$ = 3.53 from Schmutz \& Vacca (1991) in computing the average.
This revised value is based on the correct color excesses 
E$_{b-v}$ = 0.86 and E$_{B-V}$ = 1.07 (W. Schmutz, pers. comm.),
rather than the misprinted values for WR 110 that appeared in 
Schmutz \& Vacca (1991).

The temperature of the 
cool component is well-determined from both discrete-T and DEM 
fits.  This cool component is responsible for $\approx$45\%
- 55\% of the total observed (absorbed) flux in the 0.3 - 10 keV 
range and may account for as much as $\approx$80\% of the 
intrinsic (unabsorbed) flux.
The temperature of the hard component is considerably more uncertain.
If the hard emission is thermal and subject to the same absorption
as the cool component (N$_{H}$ $\approx$ 10$^{22}$ cm$^{-2}$),
then a temperature kT$_{hot}$ $\geq$ 3 keV is required. Depending
on the optically thin plasma model used, we obtain values in the 
range kT$_{hot}$ $\approx$ 3.5 - 8 keV. Values of kT$_{hot}$
somewhat below 3 keV may be possible if the hot component is
more heavily absorbed, but fit statistics based on reduced
$\chi^2$ values do not favor such models.

A possibility worth considering is whether the hard X-ray component
might be nonthermal emission. Hard X-rays with power-law spectra 
are predicted for such processes as inverse Compton scattering
(Chen \& White 1991), so we have compared models
using an optically thin thermal plasma for both the cool
and hot components with models which use a thin thermal plasma for
the cool component and a  power-law for the hot component.
The reduced $\chi^2$ values for these two cases are 
identical, with simultaneous fits of the PN spectrum and combined MOS 
spectra giving $\chi^2_{red}$ = 0.89 (96 dof) in both cases. The 
best fit photon power-law index is $q$ = 2.2 (1.9 - 2.6),
where the parentheses enclose the 90\% confidence range. 
Thus, we cannot rule out a power-law model for the hard X-ray emission
based on the X-ray spectra alone and the issues concerning  
nonthermal X-ray emission are discussed further in Section 4.5.

\subsection{Radio Properties}

Figure 6 shows the 22 GHz VLA image with a clear detection of
WR 110. The radio position (Table 2) is in excellent agreement with
the SIMBAD optical position, with a radio $-$ optical offset of
$\Delta$RA = $-$0.0007 s and $\Delta$DEC = $+$0.02$''$. As shown
in Figure 7,
the total flux S$_{\nu}^{(total)}$ increases with frequency according to a 
power law S$_{\nu}^{(total)}$ $\propto$ $\nu^{\alpha}$, where 
$\alpha$ = $+$0.64$\pm$0.10 (90\% confidence limits). This behavior
is in excellent agreement with that expected for free-free emission
from  a spherical ionized constant-velocity wind (Wright \& Barlow 1975).
The total 4.86 GHz flux S$_{4.86}$ = 1.17 $\pm$ 0.04
mJy (Table 2) is slightly larger than the
value of 0.96 $\pm$ 0.10 mJy measured with the
VLA in C-configuration in July 1980 (Bieging,
Abbott, \& Churchwell 1982). The cleaned 4.86 GHz image 
shows no significant emission ($\geq$4$\sigma$) within
30$''$ of WR 110. Thus, there is no radio evidence of background
AGNs near WR 110 that could contaminate the X-ray spectrum.
No circular polarization is detected and the 
most stringent upper limit on fractional circular polarization
($\pi_{c}$) comes from the 8.44 GHz Stokes V image. The
rms noise in the 8.44 GHz Stokes V image is 35 $\mu$Jy, 
giving a 3$\sigma$ upper limit $\pi_{c}$ $\leq$ 0.059.

The source is unresolved at the three lowest frequencies, 
as indicated by the good agreement between peak and total
fluxes (Table 2). However, at the highest frequency of 22 GHz
the peak flux is less than the total flux and the difference
is significant (5.9$\sigma$). This difference is present in
the cleaned image obtained by combining $uv$ data from all
three 22 GHz scans ($\approx$10 minutes per scan) and also in
the cleaned image of each scan. However, no significant 
difference is seen between the peak and total fluxes of 
the phase calibrator  1733$-$310. These results raise the
interesting possibility that the radio source is  partially resolved
at 22 GHz with a synthesized beam of 0.$''$6 $\times$ 0.$''$3. 
Using the current best estimates for the mass-loss parameters
given below (Sec. 3.2.1), the effective 22 GHz angular size of the
radio emitting region of an ideal spherical wind is $<$0.$''$02
at d = 1.28 kpc (Panagia \& Felli 1975). If the adopted distance
is correct, then it is unlikely that the VLA is
resolving the wind unless it is highly non-spherical. 
An alternate explanation is that two or more radio-emitting
components lie within the synthesized beam. To be marginally
resolved at 22 GHz, the angular separation between components
would need to be at least $\approx$0.$''$3, which equates
to a projected linear separation of at least $\sim$384 AU at 
d = 1.28 kpc. Given that the present evidence for partially
resolved radio emission is based on limited high-frequency
22 GHz snapshot data, we believe that confirmation will be 
necessary in higher angular resolution observations with more
complete $uv$ coverage.

\subsubsection{Mass Loss Rate}
The ionized mass loss rate for an assumed constant-velocity
wind can be estimated using the
result of  Wright \& Barlow (1975), namely 
$\mathrm{\dot{M}}$ = 
C$_{0}$v$_{\infty}$S$_{\nu}^{0.75}$d$^{1.5}$ M$_{\odot}$ 
yr$^{-1}$,
where C$_{0}$ = 0.095$\mu$/[Z$\sqrt{\gamma g \nu}$].
Here, v$_{\infty}$ (km s$^{-1}$)  is the terminal 
wind speed, S$_{\nu}$ (Jy) is the observed radio flux at
frequency $\nu$ (Hz), d (kpc) is the stellar distance,
$\mu$ is the mean atomic weight per nucleon, Z is the
rms ionic charge, $\gamma$ is the mean number of free 
electrons per nucleon, and g is the free-free Gaunt factor.
To evaluate this expression we use the highest 
signal-to-noise radio detection at  $\nu$ = 8.44 GHz,
S$_{8.44}$ = 1.77 mJy (Table 2), v$_{\infty}$ = 2100
km s$^{-1}$ (vdH01), d = 1.28 kpc (vdh01), Z = 1 
(vdH86), $\gamma$ = 0.96 (vdH86), and g = 4.78
at 8.44 GHz from the approximation for the free-free Gaunt factor
given in Eq. [8] of Abbott et al. (1986), assuming a temperature
at the radio photosphere of T = 10000 K (vdH86). These values give 
$\mathrm{\dot{M}}$ = 1.267$\mu$ $\times$ 10$^{-5}$  M$_{\odot}$ yr$^{-1}$.
Assuming $\mu$ = 3.9 for a WN star (vdH86), one obtains 
$\mathrm{\dot{M}}$ = 4.9 $\times$ 10$^{-5}$  M$_{\odot}$ yr$^{-1}$.
This mass-loss rate gives a wind luminosity
L$_{wind}$ = (1/2)$\mathrm{\dot{M}}$v$_{\infty}^2$ =
6.8 $\times$ 10$^{37}$ ergs s$^{-1}$.

\subsection{Filling Factor of X-ray Emitting Gas}

An important quantity for constraining X-ray emission models
is the average filling factor $f$ which is defined by the ratio 
$f$ = EM$_{x}$/EM$_{tot}$. Here,  EM$_{x}$ is
the volume emission measure of the X-ray emitting 
plasma and EM$_{tot}$ is the total volume emission 
measure of the wind. For radiative shock models, 
the value of $f$ provides a rough measure of the 
fraction of the wind that is undergoing instability
developments and  $f$ is also useful in
comparing the X-ray properties of WR and O-type
stars (Ignace, Oskinova, \& Foullon 2000). One
advantage of obtaining both X-ray and radio data
is that $f$ can be calculated in a straightforward manner.

EM$_{x}$ is obtained from X-ray spectral fits 
via the XSPEC normalization parameter $norm$ in MEKAL
and VMEKAL models, which is defined by 
$\int$n$_{e}$n$_{H}$dV = 10$^{14}$(4$\pi$d$^{2}$)$\cdot norm$
cm$^{-3}$. Using d = 1.28 kpc and the adopted canonical WN 
abundances with n$_{He}$ = 14.9n$_{H}$ (vdH86), we obtain the
X-ray emission measure of a helium-dominated plasma 
EM$_{x}$ = 3.12 $\times$ 10$^{59}$$\cdot norm$ cm$^{-3}$.
For an assumed constant-velocity spherical
wind,   EM$_{tot}$ is related to 
the observed radio flux via the relation given in Eq. [7]
of Abbott et al. (1986). Inserting the 8.44 GHz parameters
given above (Sec. 3.2.1) into Eq. [7] of Abbott et al.
we obtain EM$_{tot}$ = 
5.5 $\times$ 10$^{60}$$\mathcal{I}$ (R$_{*}$/R$_{\odot}$)$^{-1}$
cm$^{-3}$, where $\mathcal{I}$ = $\int$$w^{-2}dx$ is the
dimensionless wind-velocity integral that enters into the
calculation. The variable of integration is 
$x$ = R$_{*}$/r where $w$ = v/v$_{\infty}$ = 1 
and $\mathcal{I}$ = 1
for a spherical constant-velocity wind. We adopt 
the value $\mathcal{I}$ = 14 appropriate for the wind
velocity law of Abbott et al. and thereby obtain
$f$ = 4.05 $\times$ 10$^{-3}$~$\cdot$$norm$~(R$_{*}$/R$_{\odot}$).

The stellar radius is uncertain, but values in the range 
R$_{*}$ = 1.8 - 6 R$_{\odot}$ have been estimated in previous work
(Abbott et al. 1986; Hamann et al. 1995). The XSPEC
$norm$ value is also uncertain since it depends on the
abundances used to fit the X-ray spectrum. If the spectrum
is fitted with a 2T MEKAL model using canonical WN abundances 
(vdH86) then $norm$ = $norm_{1}$ $+$ $norm_{2}$ =
(1.25 $+$ 0.50) $\times$ 10$^{-5}$, where $norm_{1}$ and 
$norm_{2}$ correspond to the cool (kT$_{cool}$ = 0.55 keV) and hot
(kT$_{hot}$ $\geq$ 3 keV) components. Allowing for the 
uncertainty in R$_{*}$ we obtain 
$f$ = (1.3 - 4.1)  $\times$ 10$^{-7}$,
or an order-of-magnitude estimate $f$ $\sim$ 10$^{-7}$.
The above range in $f$ includes the contributions of
both the cool and hot plasma components. If only the cool
component is included, then $f$ decreases by 
$\approx$28\% but the order-of-magnitude result is unchanged.
The same 2T MEKAL model using solar abundances gives 
slightly larger values $f$ = (0.66 - 2.1) $\times$ 10$^{-6}$
or $f$ $\sim$ 10$^{-6}$, including contributions from both
the cool and hot components.

\section{Discussion}

The new XMM results discussed above reveal a two-component
X-ray structure for WR 110, consisting of both soft and hard
X-ray emission. We discuss possible emission mechanisms below.
The leading model for explaining the soft emission is based
on the instability-driven wind-shock picture (Sec. 4.1), while
several alternative mechanisms are examined for explaining the
hard emission (Secs. 4.2 - 4.5). 

\subsection{Instability-Driven Wind Shocks}

The instability-driven wind shock model attributes the 
the X-ray emission from single hot stars to shocks
that are distributed throughout the wind (Lucy \& White 1980; 
Lucy 1982;  Owocki et al. 1988; Feldmeier et al. 1997).
Some observational support for this model has now been obtained
in X-ray  grating spectra of the O4 supergiant $\zeta$ Puppis
(Cassinelli et al. 2001; Kahn et al. 2001). The stronger emission
lines in the $\zeta$ Pup spectrum such as Ne X are blueshifted,
consistent with predictions for line formation in an outflowing
wind (e.g. Owocki \& Cohen 2001). However, grating observations 
have so far failed to detect such Doppler shifts in other 
putatively single O-type stars (e.g. Schulz et al. 2000; Waldron \& 
Cassinelli 2001) or in the X-ray bright WC8 $+$ O7 binary
$\gamma^2$ Vel (Skinner et al. 2001). Additional X-ray grating 
observations of a larger sample of hot stars  are clearly 
needed to confirm the presence of Doppler shifts.
Another potential problem  is that
the instability-driven shock model has difficulties accounting
for the S XV line in the $\zeta$ Pup spectrum, which apparently
forms near the base of the wind where post-shock velocities 
(and the corresponding shock temperatures) are expected to be low.

The XMM-Newton data for WR 110 present a new challenge for the
instability-driven wind shock model, which predicts only soft
X-rays. Although this model might be able to explain the cool
emission component in WR 110, the hot emission component is not
predicted by current versions of this model and will thus require a
different explanation (Secs. 4.2 - 4.5). 
Assuming that the cool emission is due to instability-driven wind
shocks, we can place constraints on the range of shock velocities 
present and can also estimate the minimum radius from which the detected
X-rays emerge. 

To estimate the range of shock velocities needed to
account for the cool component, we use the 
relation for the post-shock temperature of a
strong adiabatic shock is
kT$_{s}$ = (3/16)\={m}v$_{s}^2$,
where \={m} is the mean particle mass 
in the wind and v$_{s}$  is
the shock velocity. For a helium-rich WN
wind we have \={m} = (4/3)m$_{p}$ where
m$_{p}$ is the proton mass. The observed temperature
kT$_{cool}$ = 0.55 keV gives a typical shock velocity  v$_{s}$ = 460 
km s$^{-1}$. However, the DEM distribution 
(Fig. 4) spans a range of temperatures from 
kT$_{cool}$ $\approx$ 0.3 - 0.8 keV (FWHM), so the observations
imply a range of shock velocities from 
340 - 550 km s$^{-1}$. These values are about 
a factor of two larger than predicted using
the original radiative-shock formulation of
Lucy (1982).

To estimate the minimum radius from which the detected X-rays emerge,
we assume that the wind is spherical and homogeneous. In that case,
the emergent X-rays detected in our observation must come from radii
larger than the radius R$_{\tau = 1}$(E) at which the wind becomes 
optically thin. Although  X-rays could be produced at smaller radii,
they would have escaped detection due to absorption by the overlying
wind. Using the mass-loss parameters in Sec. 3.2.1 and the WN wind 
cross-sections $\sigma_{w}$ for X-rays from Ignace et al. (2000), we obtain 
R$_{\tau = 1}$(E = 1 keV) = 1.8 $\times$ 10$^{14}$ cm 
 $\approx$12 AU.  For an assumed stellar radius R$_{*}$ =  4
R$_{\odot}$, this equates to R$_{\tau = 1}$(E = 1 keV) = 645 R$_{*}$.
By comparison, the wind absorption cross-sections are
much lower at high energies with an approximate dependence
$\sigma_{w}$ $\propto$ E$^{-2.5}$ (Fig. 1 of Ignace et al.
2000), and the hard component emission detected above $\sim$3 keV 
could thus originate at radii of less than $\sim$1 AU. 

If stellar X-rays suffered significant absorption by circumstellar
material, in addition to wind absorption, then the above calculation
would underestimate R$_{\tau = 1}$(E). However, there is no reason
to believe that significant circumstellar X-ray absorption has 
occurred for WR 110.  The value of N$_{H}$ determined from the X-ray
spectra is consistent with that inferred from the visual extinction
(Sec. 3.1.3), so the emergent X-rays have not incurred any significant
excess absorption above that seen in the optical. There is no evidence
for H or circumstellar material in the optical spectra  (Schmutz, Hamann,
\& Wessolowski 1989). For circumstellar material to have escaped 
optical detection,
it would have to be largely unionized which would be unlikely given the
high stellar effective temperature (Hamann et al. 1995). We thus believe
that the neutral hydrogen column density N$_{H}$ = 1.05 $\times$ 10$^{22}$
cm$^{-2}$ is due mainly to interstellar absorption, which is expected to
be large for WR 110 since it is viewed toward the galactic center
(galactic coordinates $l$ = 10.8, $b$ = $+$0.39).

Thus, for an ideal spherical homogeneous wind we conclude that the
emergent cool X-ray emission comes from hundreds of stellar radii
whereas the harder emission could originate much closer to the star.
If the wind is aspherical or inhomogeneous (e.g. clumpy) then
R$_{\tau = 1}$(E) depends on geometry and clump properties. In that
case, even the cool emission detected in our observation could come from
smaller radii than estimated above.

\subsection{Magnetically-Confined Wind Shocks}

The detection of a hard emission component in WR 110 is
of interest given that hot plasma has also been reported
in O-type stars such as $\theta^1$ Ori C (Schulz et al. 2000) 
and $\zeta$ Ori (Waldron \& Cassinelli 2001). The traditional
view that hot stars emit only soft weakly-absorbed
X-rays is now being replaced by a more complex picture that
also requires high-temperature plasma  originating close to
the star. It has been suggested that the hot plasma may
be magnetically confined near the base of the wind and this
has sparked renewed interest in magnetically-confined wind
shock models. Such models were originally proposed to explain
the X-ray emission of magnetic Ap-Bp stars (Babel \& Montmerle 1997a)
but were subsequently extended to young O-type stars such
as $\theta^1$ Ori C (Babel \& Montmerle 1997b). 

In this picture the ionized wind is trapped by a (dipolar)
magnetic field and is then channeled along field lines toward
the magnetic equatorial
plane  where the two streams from the separate hemispheres collide
to produce an X-ray emitting shock. A dense geometrically thin
cooling disk is predicted to form in the equatorial plane.
This process is capable of producing high-temperature plasma
and thus might be considered as a  means of 
explaining the hard emission component in WR 110.

At present, it seems quite difficult to justify the above
model for WR stars in general, and for WR 110 in particular.
There is, to our knowledge, no persuasive evidence for
magnetic activity in WR 110. This contrasts 
with O-type stars such as $\theta^1$ Ori C which show
clear rotational modulation in X-rays and H$\alpha$.
However, additional searches for evidence of magnetic 
fields would be valuable since magnetically-active regions have
been proposed as a possible explanation of the 3.766 d
optical and ultraviolet periodicity detected in the
WN 4 star EZ CMa (St.-Louis et al. 1995). Although previous
observations suggest that the X-ray emission of EZ CMa 
is variable, there is no evidence so far that the X-rays
are modulated at the 3.766 d optical period (Willis \&
Stevens 1996; Skinner, Itoh, \& Nagase 1998).

Another potential difficulty is that WR stars typically
have much larger mass-loss rates than O-type or Ap-Bp
stars and proportionally larger magnetic fields are 
required for wind confinement. 
The degree to which the wind is confined by the magnetic
field is determined by the confinement parameter 
$\Gamma$ = B$_{0}^2$R$_{*}^2$/$\mathrm{\dot{M}}$v$_{\infty}$,
where B$_{0}$ is the surface equatorial field strength
(ud-Doula \& Owocki 2002).
If $\Gamma$ $>>$ 1 then the wind is strongly confined
but if $\Gamma$ $<<$ 1 then the field is 
stretched out by the wind. Assuming R$_{*}$ = 4 R$_{\odot}$ 
for WR 110 and using the adopted mass-loss parameters (Sec. 3.2.1)
gives B$_{0}$ = 2.9 $\sqrt{\Gamma}$ kG.
Thus, surface fields of several kG are required
for confinement. Electrons trapped in such a strong field
should emit nonthermal radio emission as recognized by
Babel \&  Montmerle (1997a), but such emission could be
masked by heavy wind absorption in WR stars.

Magnetic fields of several kG would also imply much higher 
X-ray luminosities from a magnetically-confined wind
shock than derived from the XMM spectra. The predicted
luminosity for a marginally confined wind in WR 110  
using B$_{0}$ = 2.9 kG is at least L$_{x}$ $\approx$
10$^{36.6}$ ergs s$^{-1}$ (eq. [10] of Babel \& Montmerle
1997a). This value is four orders of magnitude larger 
than observed, which is a large mismatch. We also 
note that very efficient conversion of wind kinetic energy
to thermal X-rays was required for this model to successfully
explain the X-ray emission of the Ap star IQ Aur (Babel \&
Montmerle 1997a). In contrast, the mechanism responsible for
the X-ray emission in WR 110 is very inefficient since
L$_{x}$/L$_{wind}$ $\sim$ 10$^{-5}$.

Thus, in its current form, there are several difficulties that
need to be addressed in order to extrapolate the 
magnetically-confined wind shock model to WR stars. 
Obviously, persuasive evidence for magnetic activity is
the main missing link and further observational work is needed
to search for periodic variability in emission lines and 
the continuum at all wavelengths (including X-rays) as well
as nonthermal radio emission. Further extensions of the
theory are also needed to test it in the high mass-loss 
WR regime. The distortion of the magnetic field by the
wind is of particular importance in the case of WR stars,
and the effects of dipole magnetic fields on line-driven outflows 
have been incorporated into recent numerical magnetohydrodynamic
simulations (ud-Doula \& Owocki 2002).

\subsection{Wind Accretion Shocks}

Given the potential difficulties with magnetic wind
shock models for WR stars, we consider other alternatives
for explaining the hard X-ray emission.
It is conceivable  that the hard X-rays are due either to
gravitational accretion of the WR wind onto an optically
faint companion which has so far escaped detection, or
to the WR wind shocking onto such a companion.

Gravitational accretion onto
an optically faint neutron star companion is one
possibility that is consistent with evolutionary
scenarios for massive binaries (van den Heuvel 1976).
However, the expected
X-ray luminosity from accretion of the WR wind onto a
neutron star is much larger than observed
and we believe this option is unlikely. Specifically,
for an assumed neutron star mass  M$_{ns}$ = 1.4 M$_{\odot}$
and  M$_{WR}$ = 7.7 M$_{\odot}$ (Hamann et al. 1995),
the results of Davidson \& Ostriker (1973) give a predicted
unabsorbed accretion luminosity of at least  L$_{x,acc}$   
$\sim$ 10$^{37}$ ergs s$^{-1}$. This value is several 
orders of magnitude larger than the intrinsic luminosity
of WR 110 obtained from spectral fits (Table 1).

Gravitational accretion onto a faint normal (nondegenerate)
stellar companion is also possible, but again seems unlikely.
The gravitational radius is R$_{G}$ = 2GM$_{comp}$/v$_{w}^{2}$ 
where G is the gravitational constant, M$_{comp}$ is the 
mass of the putative companion, and v$_{w}$ is the WR wind
velocity. Using v$_{w}$ = 2100 km s$^{-1}$ gives 
R$_{G}$/R$_{\odot}$ = 0.38 (M$_{comp}$/M$_{\odot}$)/v$_{1000}^{2}$,
where  v$_{1000}$ is the WR wind 
velocity in units of 1000 km s$^{-1}$.
Assuming a solar-like companion with M$_{comp}$ $\approx$ M$_{\odot}$
and R$_{comp}$ $\approx$ R$_{\odot}$ we obtain  R$_{G}$ = 0.09 R$_{\odot}$.
Thus,  R$_{G}$ $<<$  R$_{comp}$ and gravitational accretion is
ineffective. We thus consider the more likely case below in which
the WR wind collides directly with the surface of the 
putative companion.

\subsection{Colliding Wind Shocks}

Although the temperature of the hot component is not tightly
constrained, spectral fits with optically thin plasma models
are consistent with values as high as kT$_{hot}$ $\approx$ 8
keV. Such high temperatures could be produced if the WR wind
overwhelms that of a lesser companion, forming an adiabatic
shock at or near the companion surface. In that case the
temperature relation for a strong adiabatic shock gives
kT$_{s}$ = (3/16)\={m}v$_{s}^2$ $\approx$ 11 keV, where 
we have assumed that the WR wind has reached terminal speed
and thus v$_{s}$ $\approx$ v$_{\infty}$ = 2100 km s$^{-1}$.
Assuming that the hot plasma detected in WR 110 is free-free
emission from such an adiabatic shock, we can place constraints
on the companion separation $a$ using two different approaches
based on timescales and X-ray luminosity.

The relevant dynamical timescale is t$_{d}$ =
$a$/v$_{\infty}$, where v$_{\infty}$ is the terminal
speed of the WR wind. If the high-temperature plasma is
free-free emission from a strong adiabatic shock at (or near)
the companion surface, then the cooling time t$_{cool}$ 
of the plasma behind the shock must be greater than
t$_{d}$ while the opposite is true for the electron-ion
equilibration time t$_{ei}$. Thus
t$_{ei}$ $<$ t$_{d}$ $<$ t$_{cool}$.

In the case of a pure helium wind we can write
t$_{ei}$ = 8.38 T$_{s}^{3/2}$ n$_{He}^{-1}$ (Spitzer 1962) and
t$_{cool}$ = (3/2) (n$_{He}$ + n$_{e}$) kT$_{s}$  $\epsilon_{ff}^{-1}$
 = 5.41 $\times$ 10$^{10}$ T$_{s}^{1/2}$ n$_{He}^{-1}$.
Here, we have used the explicit expression for  free-free
emissivity $\epsilon_{ff}$ from Allen (1973) assuming a Gaunt factor
of unity. Using the standard expressions for the post-shock
number density n$_{He}$ and temperature T$_{s}$ and the 
assumption of a spherical wind, the above
inequality becomes 
0.68 $\mathrm{\dot{M}}_5$ v$_{1000}^{-3}$ $<$ $a_{AU}$ $<$ 144.6 
$\mathrm{\dot{M}}_5$ v$_{1000}^{-5}$,
where $\mathrm{\dot{M}_{5}}$ is the WR mass loss
rate in units of 10$^{-5}$ M$_{\odot}$ yr$^{-1}$,
v$_{1000}$ is the terminal wind velocity in units of
1000 km s$^{-1}$, and $a_{AU}$ is the companion separation in AU.
Substituting the WR 110 mass-loss parameters (Sec. 3.2.1) 
gives 0.36 $<$ $a_{AU}$ $<$ 17.3.

A more stringent constraint on $a$ is obtained by 
equating the intrinsic X-ray luminosity of the hard
component with the adiabatic shock luminosity
L$_{x,s}$. The intrinsic (unabsorbed) flux of the hot
component determined from the thermal thin plasma models of 
the X-ray spectra is
F$_{x,hot}$(0.3 - 10 keV) = 4 $\times$ 10$^{-13}$ ergs cm$^{-2}$,
which gives L$_{x,s}$ =  7.8 $\times$ 10$^{31}$ ergs s$^{-1}$.
This value is of course a lower limit since our flux 
measurement is restricted to a specific energy range.
We assume that the strong WR wind overwhelms that of the
faint companion, in which case the companion can be 
approximated by a hard sphere. Inserting the above
value of L$_{x,s}$ and the WR 110 mass-loss parameters into
eq. [81] of Usov (1992) gives $a_{AU}$ = 
0.28(R$_{comp}$/R$_{\odot})^{0.75}$. 

If R$_{comp}$ =  1.4 R$_{\odot}$ then the above luminosity
constraint gives $a$ = 0.36 AU, which is consistent
with the lower limit  derived from the timescale argument.
If the putative companion were a main-sequence star then 
the minimum radius R$_{comp}$ =  1.4 R$_{\odot}$ would
correspond to a late A spectral type. However, there is
considerable leeway in the spectral type since the luminosity
class is unknown and larger values of R$_{comp}$ are required
if $a$ $>$ 0.36 AU. At the minimum separation $a$ = 0.36 AU,
some absorption of
the hard X-rays could occur, but as noted above the amount
of absorption will depend on the symmetry and homogeneity
properties of the wind.

\subsection{Nonthermal X-ray Emission}

As noted in Sec. 3.1.3, we cannot formally rule out a  
power-law model for the hard X-ray emission based on X-ray
spectral fits alone. We thus comment briefly on nonthermal
emission models. In the model proposed by  Chen \& White (1991,
hereafter CW91), hard X-ray photons can be produced in hot
stars by inverse Compton scattering of stellar UV
photons off of relativistic shock-accelerated particles
in the wind. This model was applied to OB supergiants by CW91
but it may have broader relevance to all hot stars that
are subject to line-driven instabilities. However,
bremsstrahlung may be as important
as inverse Compton scattering in dense WR winds (CW91).

The inverse Compton model makes some predictions that can
be compared with  observations. The X-rays are not expected
to show large amplitude variations and this is consistent
with the temporal behavior of WR 110, at least for the 
relatively short seven hour observation analyzed here.
The predicted photon power-law 
index in the hard X-ray band is $q$  =  1.5,
whereas the value inferred from the EPIC spectra for WR 110 
is $q$ = 2.2 (1.9 - 2.6). Given the relatively low signal-to-noise
ratio in the  EPIC spectra above $\sim$3 keV, this difference
may not be problematic. The model also provides an expression 
for evaluating the hard-band X-ray luminosity but it is quite 
sensitive to the assumed wind temperature and mass-loss 
parameters (eqs. [31] - [32] of CW91). Evaluation of the 
expression for the hard-band luminosity also requires a value
for the stellar magnetic field strength, which is not known
for WR 110 or for WR stars in general.

The hard X-rays from inverse Compton scattering in the CW91
model are predicted to originate within 10 R$_{*}$ of the 
star where the stellar UV photon number density is high.
This is a potential problem for WR stars since their
winds could be optically thick even to harder X-rays of several
keV this close to the star (Sec. 4.1). However, this problem 
is mitigated if the wind is clumped.

The inverse Compton model does not predict a coincidence 
between nonthermal X-rays and nonthermal radio emission.
Thus, the absence of nonthermal radio signatures in WR 110
is not necessarily in conflict with an inverse Compton
origin for the hard X-rays. If radio synchrotron emission
is formed within the wind (White 1985), it could be absorbed.
Using the adopted mass-loss parameters 
for WR 110, we estimate R$_{\tau=1}$ $\approx$ 3380 R$_{\odot}$
at $\nu$ = 8.44 GHz, or R$_{\tau=1}$ $\approx$ 845  R$_{*}$ 
assuming R$_{*}$ $\approx$ 4 R$_{\odot}$. Thus, nonthermal
radio emission formed even at hundreds of stellar radii
could escape detection.

Although an interpretation of the hard X-rays in terms of the 
inverse Compton model does not appear to conflict with our data,
there are some concerns about the relevance of this model to
WR 110.  Without specific knowledge of wind clumping properties,
it is not clear that hard X-rays (kT $\geq$ 3 keV) produced 
within $\sim$10 R$_{*}$ would penetrate the dense overlying
wind and be detected. Furthermore, direct evidence for relativistic
particles in the wind is lacking since the VLA data show no 
obvious signs of nonthermal radio emission (e.g. synchrotron
emission). To our knowledge, there have 
been no confirmed detections of nonthermal X-ray emission in
any WR star to date.  However, hard {\em thermal} X-rays 
accompanied by Fe K emission lines have been detected in some
well-studied WR $+$ OB binaries such as  
WR 140 (Koyama et al. 1994) and $\gamma^2$ Vel (Skinner et al. 2001). 
Considering the above factors, we believe that the hard X-rays in
WR 110 are more likely to be of thermal origin rather than nonthermal.

\section{Summary and Outlook}

XMM-Newton provides the largest effective area ever achieved
in an imaging X-ray telescope  and we have utilized this new
capability to obtain the first high-quality X-ray spectrum of
a Wolf-Rayet star with no known companion (WR 110). We have also  
acquired the first multifrequency radio observations of  WR 110 which
clearly show that its spectral energy distribution is in good
agreement with that expected for free-free wind emission.

The new X-ray data yield some surprises and raise several new
questions about the origin of X-ray emission in WR stars. We 
have shown that the X-ray emission is dominated by a cool component
with a characteristic temperature kT$_{cool}$ 
$\approx$ 0.5 keV ($\approx$6 MK),
but that a hard component is also present. The presence of 
hard emission is not anticipated on the basis of current
instability-driven wind shock models for single hot stars, and
we have argued that the hard emission could be due to the WR
wind shocking onto an as yet undetected close companion.
However, until compelling evidence for binarity is found
we cannot exclude the possibility that the hard emission is
intrinsic to WR 110.

Further observational and numerical work will be needed to
address the following issues:

\begin{enumerate}

\item Numerical simulations of instability-driven wind
shocks in WR stars are needed to produce synthetic spectra
that can be directly compared with recent X-ray observations. 
One question to be addressed is whether instability-driven
shocks can account for higher temperature spectral lines such as
S XV (T$_{max}$ $\sim$ 16 MK) that are seen in WR 110 and in
other putatively single O-type stars such as $\zeta$ Pup and  
$\zeta$ Ori. Such simulations will also need to determine if
radiative shocks  can persist to hundreds of stellar
radii in WR winds, in analogy with studies for O-type stars
recently undertaken by Runacres \& Owocki (2002). If not,
then a clumped wind may be needed to reconcile the cool emission
detected in WR 110 with the conventional wind shock picture.

\item Further observational work is needed to determine if the
hard X-ray emission detected in WR 110 is a common feature
of  ``single'' WR stars, or an anomaly. Good-quality X-ray spectra
of a larger sample of putatively single WR stars are needed to 
answer this question.

\item Further searches for evidence of binarity in WR 110 and
other X-ray emitting ``single'' WR stars are needed. We have
argued that a close stellar companion could explain the hard
emission in WR 110, and binarity is a good bet since 
38\% of presently catalogued WR stars are {\em known}
binaries (vdH01).  Given the difficulty of detecting faint companions
around high-luminosity WR stars (L$_{bol}$ $\sim$ 
10$^{5}$ L$_{\odot}$), the true binary fraction is undoubtedly higher.
If close companions are ultimately detected around X-ray emitting 
WR stars that are now thought to be single, then legitimate single
WR stars may turn out to be much fainter in X-rays than currently
believed. 

\item If sensitive searches for binarity in ``single'' WR stars
yield negative results, then the instability-driven wind shock 
paradigm may need to be revised. Further work on magnetically-confined
wind shock models is needed to assess their relevance to WR stars
in the high mass-loss regime where the wind stretches out the field.
Additional hard X-ray and $\gamma$-ray observations of WR stars 
should also be undertaken to search for high-energy emission in
the MeV to GeV range that could arise from nonthermal processes
such as inverse Compton scattering.

\end{enumerate}

%\placetable{tbl-1}
%\placetable{tbl-2}
%\placefigure{fig3}

\acknowledgments

This work was supported by NASA grant NAG5-10362. 
SZh acknowledges the support of  NASA grant NAG5-8236.
Research at PSI is supported by grant  2100-049343
from the Swiss NSF.
This work was based on observations obtained with 
XMM-Newton, an ESA science mission with instruments and
contributions directly funded by ESA member states
and the USA (NASA). We thank members of the XMM-Newton,
VLA, and HEASARC (NASA/GSFC) support teams for their
assistance. This research has made use of the SIMBAD
astronomical database operated by CDS at Strasbourg,
France. 

\clearpage
% ---------------------------
% TABLE1.TEX 
\begin{deluxetable}{ll}
%\tablewidth{33pc}
\tablewidth{0pc}
\tablecaption{X-ray Properties of WR110\tablenotemark{a} \label{tbl-1}}
\tablehead{
\colhead{Parameter}      &
\colhead{Value  }       
}
\startdata
PN Count Rate (c s$^{-1}$)                & 0.085 \nl
MOS Count Rate (c s$^{-1}$)               & 0.039 (M1), 0.041 (M2) \nl             
N$_{H}$ (10$^{22}$ cm$^{-2}$)             & 1.05 $\pm$ 0.18   \nl
kT (keV)                                  & 0.55 $\pm$ 0.07 $+$ hot \nl
Flux (10$^{-12}$ ergs cm$^{-2}$ s$^{-1}$) & 0.37 (2.1)          \nl
L$_{x}$ (10$^{32}$ ergs s$^{-1}$)         & 0.72 (4.1)         \nl
\tablenotetext{a} {The X-ray position obtained by averaging the 
results of all three EPIC cameras is RA(2000) = 18 h 07 m 56.95 s,
DEC(2000) = $-$19$^{\circ}$ 23$'$ 57.4$''$. The column density
(N$_{H}$) and X-ray temperature (kT) were derived from best-fit
single-absorber MEKAL and C6PVMKL models using the canonical 
WN abundances given in vdH86. The temperature of the hot 
component is uncertain, but if N$_{H}$ is the same for the 
hot and cool components then kT$_{hot}$ $\geq$ 3 keV.
Flux and L$_{x}$
are the observed (absorbed) values in the 0.3 - 10 keV range,
followed in parentheses by intrinsic (unabsorbed) values.
A distance of 1.28 kpc is assumed.}
\enddata
\end{deluxetable}
\clearpage
% TABLE2.TEX
\begin{deluxetable}{llcccc}
\tablewidth{0pc}
\tablecaption{VLA Observations of WR110\tablenotemark{a} }
\tablehead{
\colhead{Frequency} &
\colhead{Beam FWHM } &
\colhead{Duration} &
\colhead{RMS Noise } &
\colhead{Peak Flux} &
\colhead{Total Flux} \\
\colhead{(GHz)         } &
\colhead{(arcsec)} &
\colhead{(min.)} &
\colhead{($\mu$Jy/beam)} &
\colhead{(mJy/beam)} &
\colhead{(mJy)}  \\
}
\startdata
4.86  & 2.9 $\times$ 1.4 & 23 & 41  &  1.22 & 1.17  \nl
8.44  & 1.5 $\times$ 0.8 & 22 & 35  &  1.77 & 1.77  \nl
14.94 & 0.8 $\times$ 0.4 & 31 & 100 &  2.44 & 2.46  \nl
22.46 & 0.6 $\times$ 0.3 & 31 & 150 &  2.20 & 3.09  \nl
\tablenotetext{a}{All data were obtained in B configuration
on 1999 Dec 26 from 
1855 - 2152 UT. Observations were obtained at each frequency in
two orthogonal polarization channels, each with a bandwidth of 50 MHz.
Fluxes and beam sizes are from cleaned Stokes I maps. Primary
flux calibrator was 3C48. The radio position of WR110 measured
from 8.44 GHz maps is RA(2000) = 18~h 07~m 56.959~s, DEC(2000) =
$-$19$^{\circ}$ 23$'$ 56.85$''$.}
\enddata
\end{deluxetable}
\clearpage

\clearpage

\figcaption{Smoothed EPIC-PN image of the region within $\approx$3$'$ of
WR 110 (arrow). $Left$: Broad-band (0.3 - 10 keV).~
$Right$: Hard-band (6 - 10 keV).
\label{fig1}}

\figcaption{Broad-band (0.3 - 10 keV) background-subtracted 
EPIC-MOS light curve of WR 110 obtained by summing data from
the MOS-1 and MOS-2 detectors. Error bars are 1$\sigma$
and the binsize is 512 s.  Solid line is the best-fit model
for an assumed constant count-rate source. \label{fig2} }

\figcaption{Background-subtracted EPIC-PN spectrum of WR 110
rebinned for display.
\label{fig3} }

\figcaption{Differential emission measure (DEM) model of WR 110
based on a fit of the background-subtracted EPIC-PN spectrum 
using the Chebyshev 
polynomial algorithm C6PVMKL. Solid line uses canonical WN
abundances (vdH86) and dashed line shows the same model using
solar abundances (Anders \& Grevesse 1989).  
\label{fig4} }

\figcaption{Unfolded rebinned EPIC-PN spectrum of WR 110  (solid
line), overlaid with a cool optically thin VMEKAL plasma model 
with kT = 0.55 keV and N$_{H}$ = 1.1 $\times$ 10$^{22}$
cm$^{-2}$ (dashed line). Abundances of Fe, Si, and S have
been varied to fit the emission line profiles. Note the
significant hard excess above $\sim$3 keV.
\label{fig5} }

\figcaption{Cleaned 22 GHz VLA image of WR 110 obtained on
26 Dec 1999. The image is based on $\approx$30 min of on-source
time acquired in three scans of 10 min each. The deconvolved
beam size (lower right) is FWHM = 0.$''$6 $\times$ 0.$''$3.
The $rms$ noise (1$\sigma$) is 0.15 mJy and contour levels
are ($-$3, 3, 6, 9, 12, 15, 18, 21)$\sigma$.
\label{fig6} }

\figcaption{The radio spectral energy distribution of WR 110
based on total fluxes measured in near-simultaneous observations
at 4.86, 8.44, 14.94, and 22.46 GHz on 26 Dec 1999. The solid
line is a best-fit power-law model, which gives a spectral
index $\alpha$ = $+$0.64 $\pm$ 0.10 (90\% confidence).
\label{fig7} }

%%\end{document}

\clearpage

\begin{figure}
\figurenum{1}
\epsscale{0.6}
\plotone{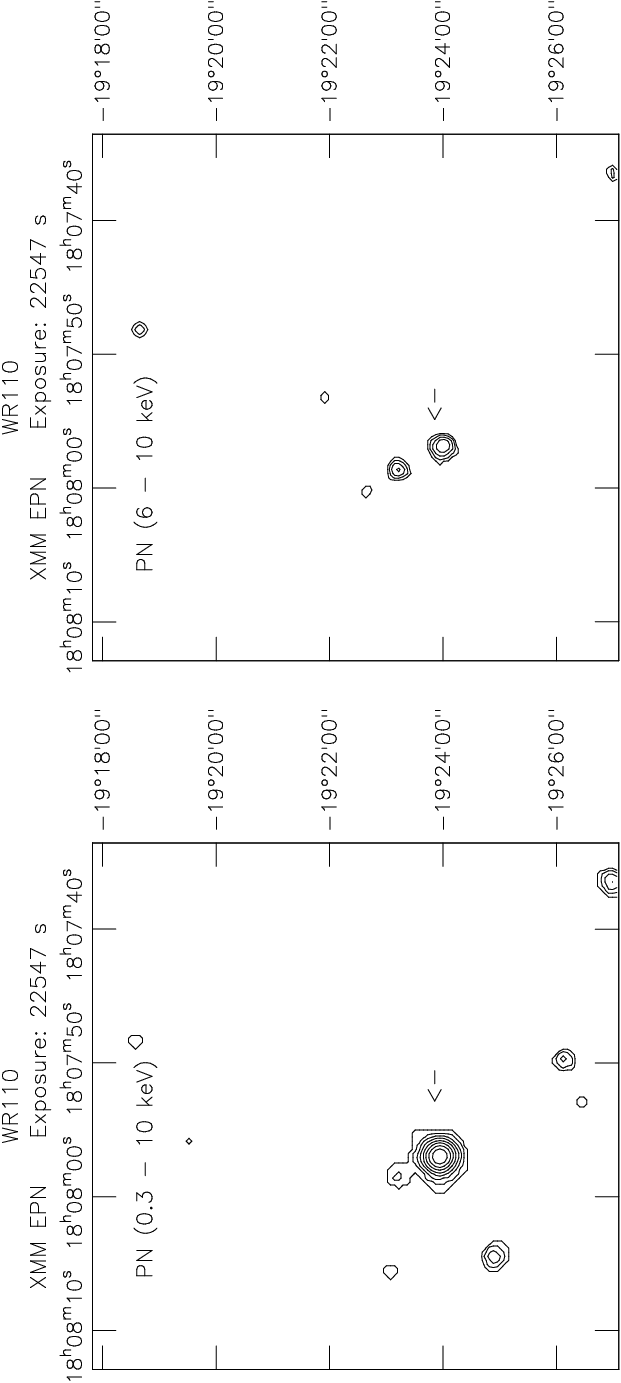}
%\plotfiddle{fig1.ps}{7 in}{90}{1.0}{1.0}{0}{36}
\caption{}

\end{figure}
 
\clearpage

\begin{figure}
\figurenum{2}
\epsscale{1.0}
\plotone{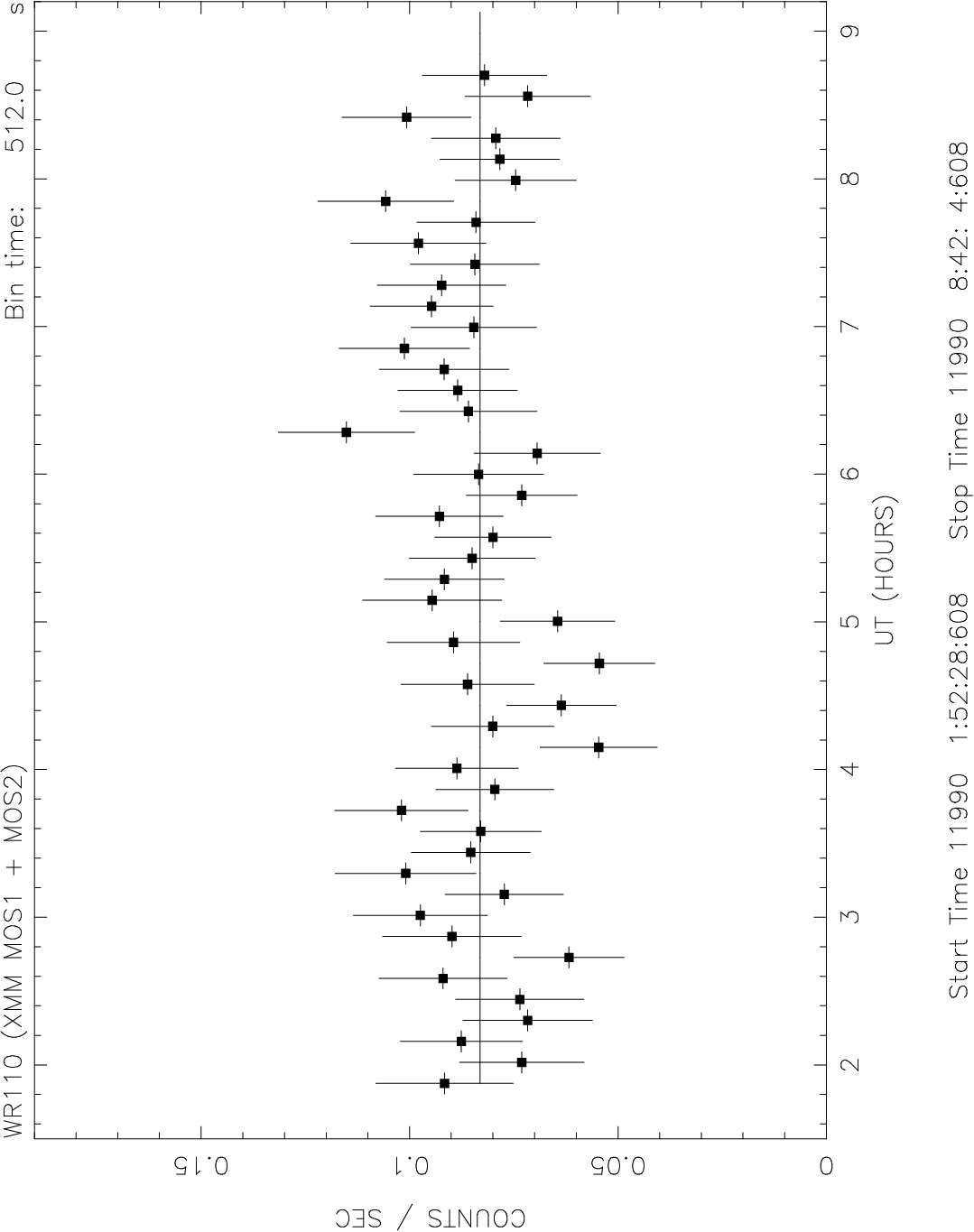}
%\plotfiddle{fig2.ps}{7 in}{270}{0.7}{0.7}{144}{144}
\caption{}
\end{figure}

\clearpage

\begin{figure}
\figurenum{3}
\epsscale{0.9}
\plotone{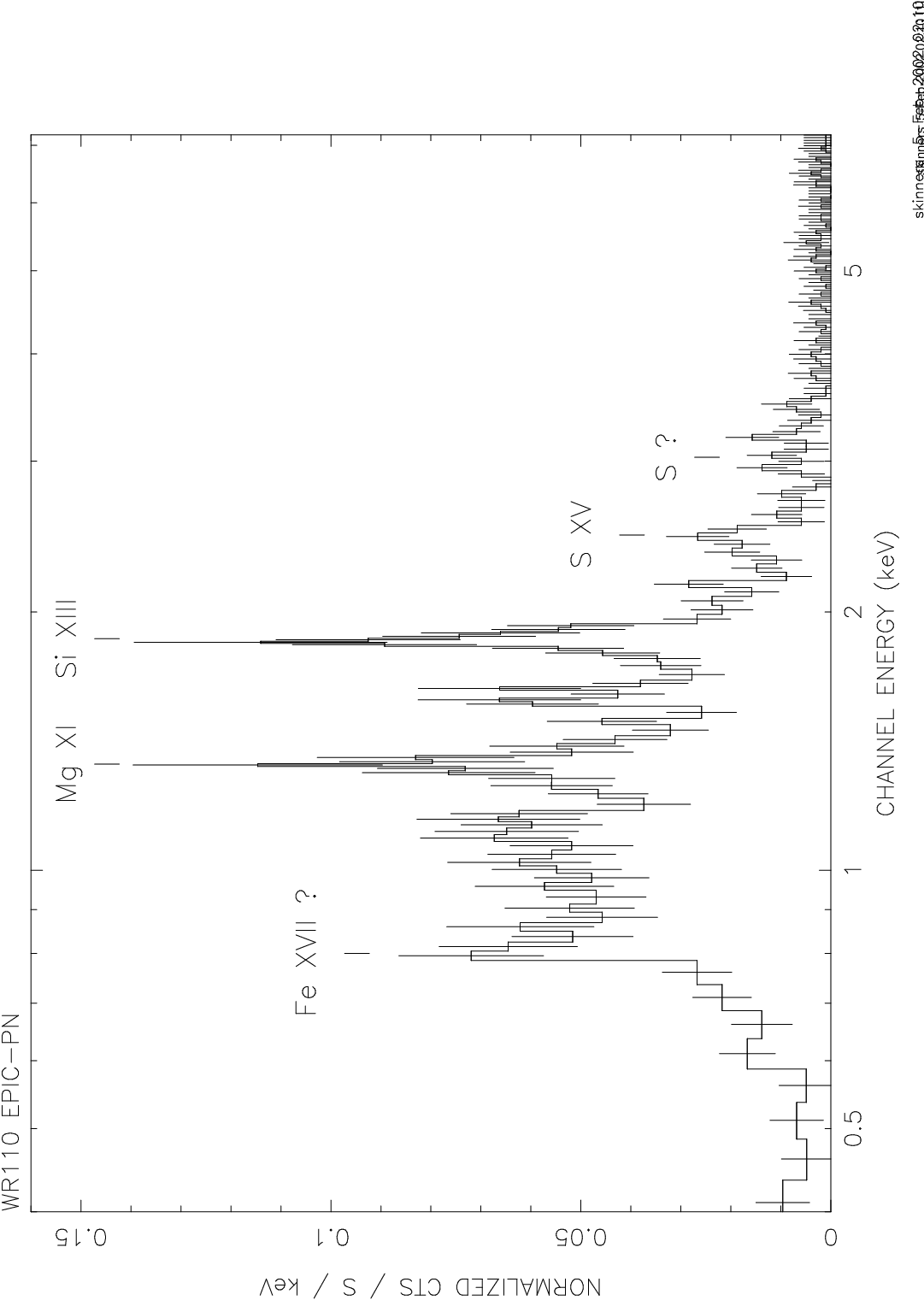}
%\plotfiddle{fig3.ps}{7 in}{90}{1.0}{1.0}{0}{36}
\caption{}
\end{figure}

\clearpage

\begin{figure}
\figurenum{4}
\epsscale{0.9}
\plotone{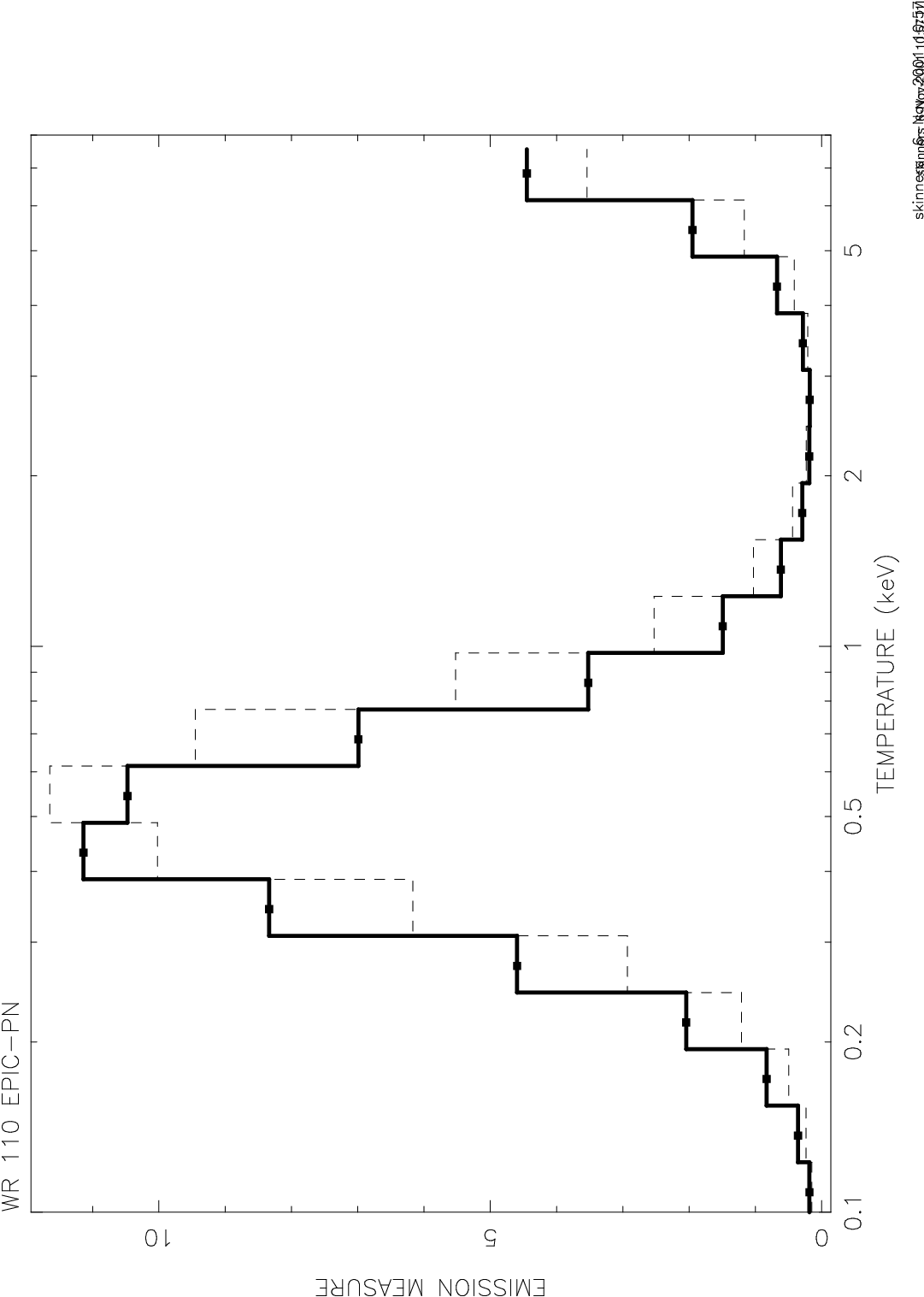}
%\plotfiddle{fig4.ps}{7 in}{90}{1.0}{1.0}{0}{36}
\caption{}
\end{figure}

\clearpage

\begin{figure}
\figurenum{5}
\epsscale{0.9}
\plotone{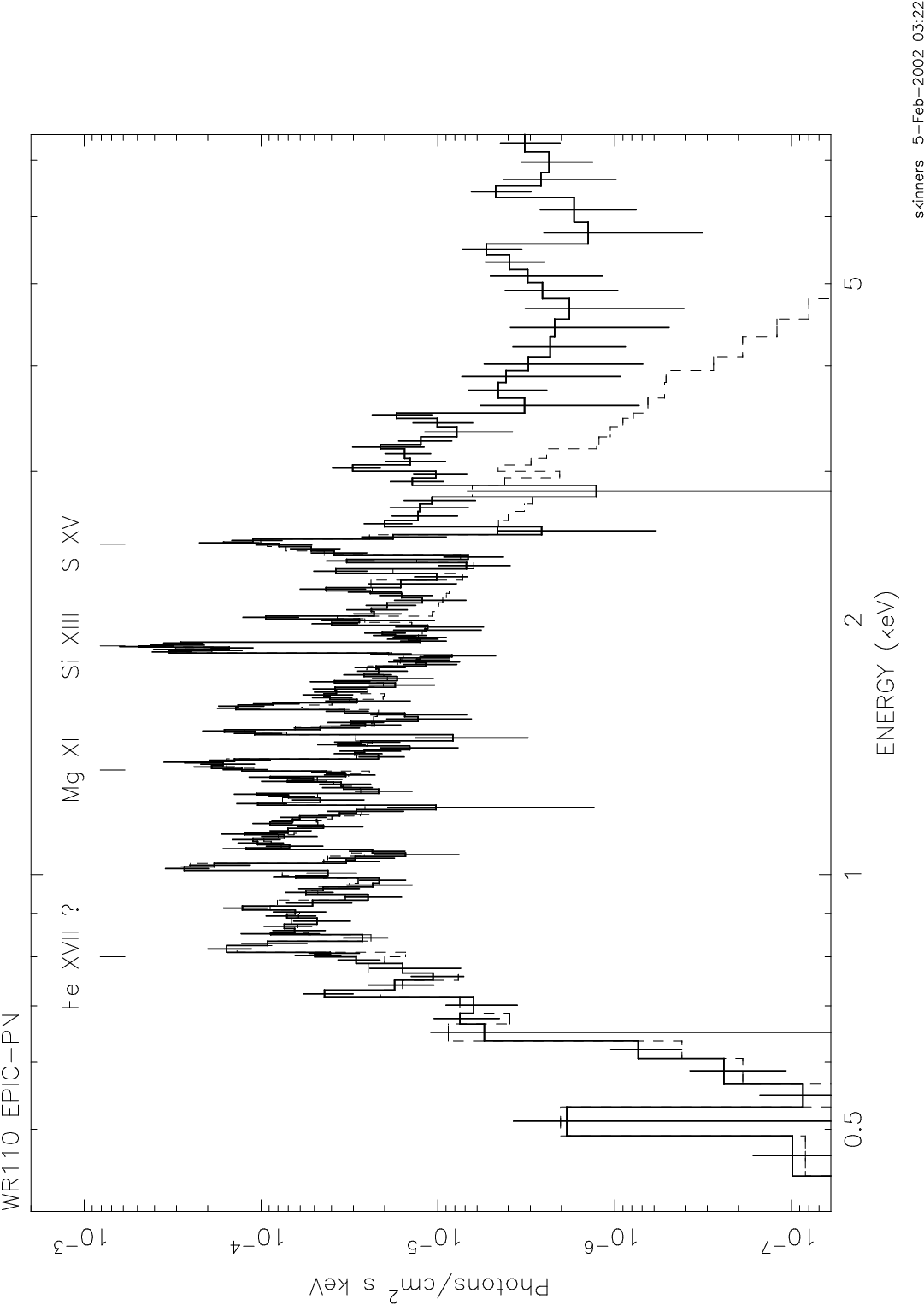}
%\plotfiddle{fig3.ps}{7 in}{90}{1.0}{1.0}{0}{36}
\caption{}
\end{figure}

\clearpage

\begin{figure}
\figurenum{6}
\epsscale{1.0}
\plotone{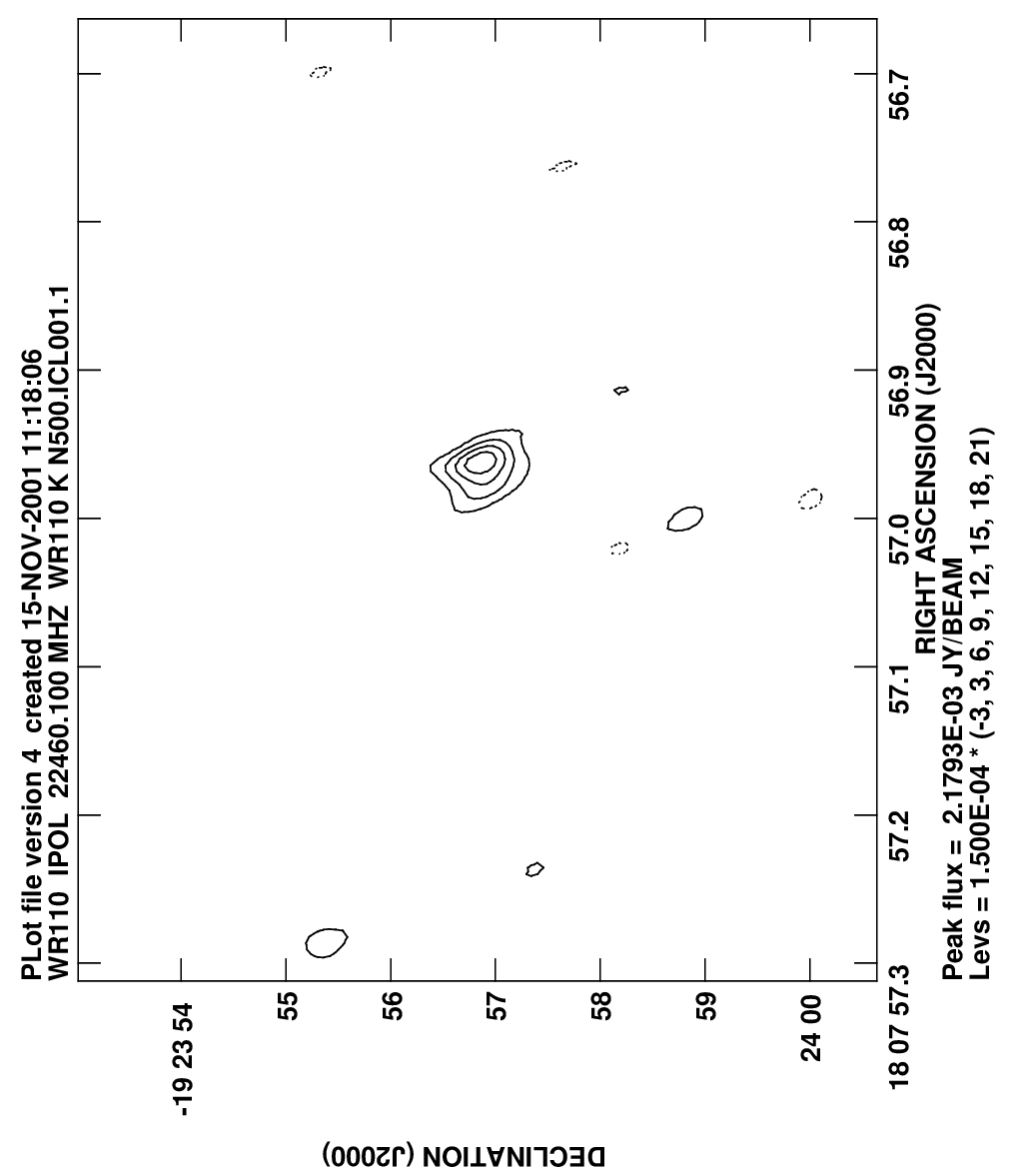}
%\plotfiddle{fig3.ps}{7 in}{90}{1.0}{1.0}{0}{36}
\caption{}
\end{figure}

\clearpage

\begin{figure}
\figurenum{7}
\epsscale{0.9}
\plotone{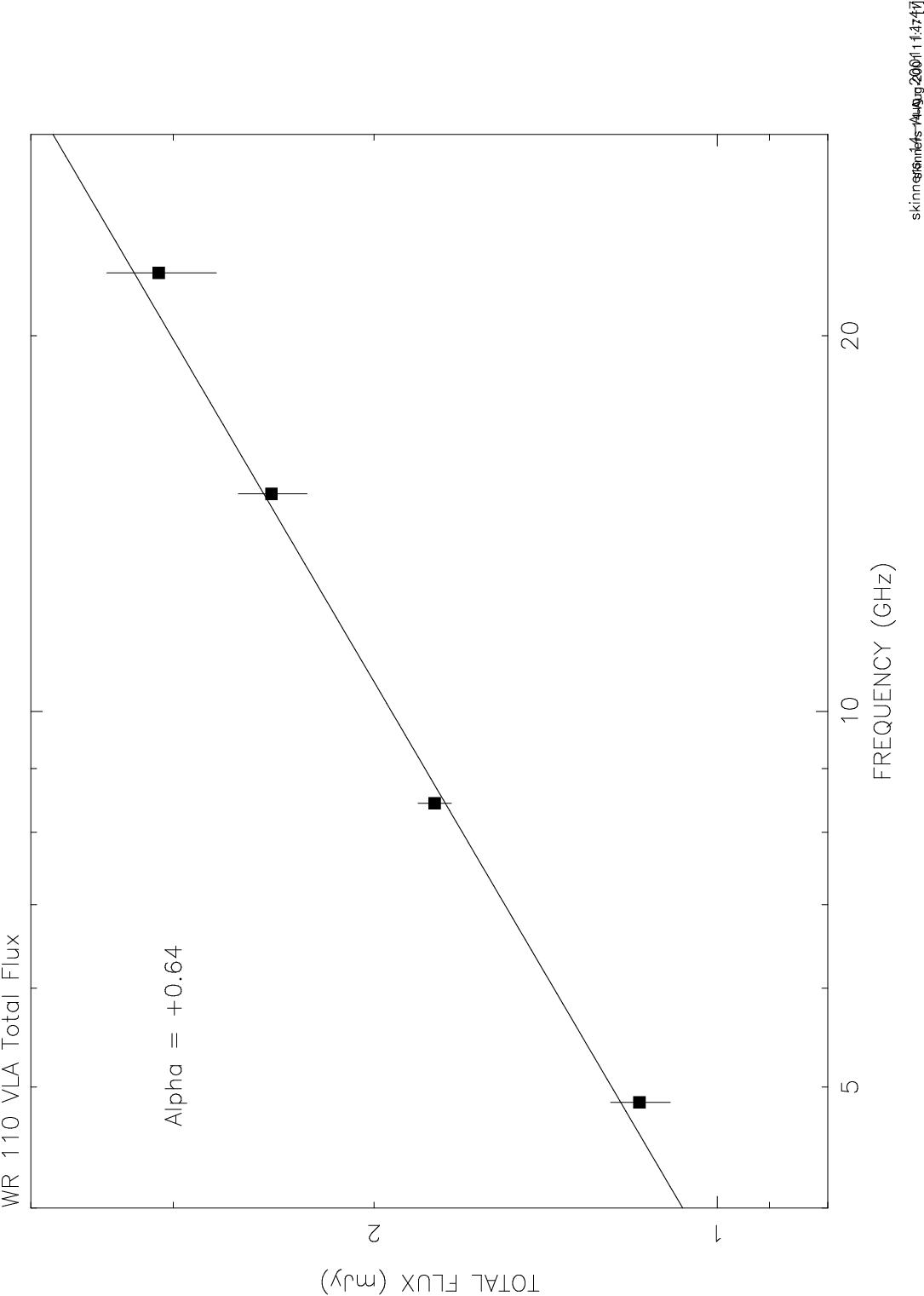}
%\plotfiddle{fig3.ps}{7 in}{90}{1.0}{1.0}{0}{36}
\caption{}
\end{figure}


\begin{thebibliography}{}
\bibitem[]{} Abbott, D.C., Bieging, J.H., Churchwell, E., 
\& Torres, A.V. 1986, \apj, 303, 239
\bibitem[]{} Allen, C.W. 1973, Astrophysical Quantities
(Athlone Press)
\bibitem[]{} Anders, E., \& Grevesse, N. 1989, \gca, 53, 197
\bibitem[]{} Arnaud, K.A. 1996, in Astronomical Data Analysis 
Software and Systems V, eds. G. Jacoby \&  J. Barnes, 
(San Francisco: ASP), 101, 17 
\bibitem[]{} Babel, J. \& Montmerle, T., 1997a, \aap, 323, 121
\bibitem[]{} Babel, J. \& Montmerle, T., 1997b, \apj, 485, L29
\bibitem[]{} Bergh\"{o}fer, T.W., Schmitt, J.H.M.M., Danner, R., 
\& Cassinelli, J.P., 1997, \aap, 322, 167
\bibitem[]{} Bieging, J.H., Abbott, D.C., \& Churchwell, E.B.,
1982, \apj, 263, 207
\bibitem[]{} Cassinelli, J.P., Miller, N.A., Waldron, W.A., 
MacFarlane, J.J., \& Cohen, D.H., 2001, \apj, 554, L55
\bibitem[]{} Chen, W. \& White, R.L., 1991, \apj, 366, 512 (CW91)
\bibitem[]{} Davidson, K. \& Ostriker, J.P., 1973, \apj, 179, 585
\bibitem[]{} Dulk, G.A., 1985, \araa, 23, 169
\bibitem[]{} Feldmeier, A., Kudritzki. R.-P., Palsa, R.
Pauldrach, A.W.A., \& Puls, J., 1997, \aap, 320, 899
\bibitem[]{} Gayley, K.G. \& Owocki, S.P., 1995, \apj, 446, 801
\bibitem[]{} Gorenstein, P. 1975, \apj, 198, 95
\bibitem[]{} Hamann, W.-R., Koesterke, L., \& Wessolowski, U., 1995,
\aap, 299, 151
\bibitem[]{} Ignace, R., Oskinova, L.M., \& Foullon, C., 2000,
\mnras, 318, 214
\bibitem[]{} Jansen, F. et al., 2001, \aap, 365, L1
\bibitem[]{} Kahn, S.M. et al., 2001, \aap, 365, L312
\bibitem[]{} Koyama, K., Maeda, Y., Tsuru, T., Nagase, F., 
\& Skinner, S., 1994, \pasj, 46, L93
\bibitem[]{} Lemen, J.R., Mewe, R., Schrijver, C.J., \& Fludra, A., 1989,
\apj, 341, 484
\bibitem[]{} Lucy, L.B., 1982, \apj, 255, 286
\bibitem[]{} Lucy, L.B. \& White, R.L., 1980, \apj, 241, 300
\bibitem[]{} Luo, D., McCray, R., \& MacLow, M-M., 1990, \apj, 362, 267
\bibitem[]{} Mewe, R., Gronenschild, E.H.B.M., \& van den Oord, G.H.J., 
1985, \aaps, 62, 197
%%\bibitem[]{} Mewe, R., Kaastra, J.S., \& Liedahl, D.A. 1995,
%%Legacy, 6, 16
\bibitem[]{} Morrison, R. \& McCammon, D. 1983, \apj, 270, 119
\bibitem[]{} Owocki, S.P., Castor, J.I., \& Rybicki, G.B., 1988,
\apj, 335, 914
\bibitem[]{} Owocki, S.P. \& Cohen, D.H. 2001, \apj, 559, 1108
\bibitem[]{} Panagia, N. \& Felli, M., 1975, \aap, 39, 1
\bibitem[]{} Pittard, J.M. et al., 2001, \aap, submitted
\bibitem[]{} Prilutskii, O., Usov, V.V., 1976, SvA-AJ, 20,2
%%\bibitem[]{} Raymond, J.C., Smith, B. W., 1977, \apjs, 35, 419
\bibitem[]{} Runacres, M.C. \& Owocki, S.P., 2002, \aap, 381, 1015
\bibitem[]{} Schmutz, W., Hamann, W.-R., \& Wessolowski, U., 1989,
\aap, 210, 236
\bibitem[]{} Schmutz, W. \& Vacca, W.D., 1991, \aaps, 89, 259
\bibitem[]{} Schmutz, W. et al., 2002, in New Visions of the 
X-ray Universe, in prep.
\bibitem[]{} Schulz, N.S., Canizares, C.R., Huenemoerder, D., \&
Lee, J.C., 2000, \apj, 545, L135
\bibitem[]{} Skinner, S.L., G\"{u}del, M., Schmutz, W., \& Stevens, I.R.,
2001, \apj, 558, L113
\bibitem[]{} Skinner, S.L., Itoh, M., \& Nagase, F. 1998,
New Astron., 3, 37
\bibitem[]{} Spitzer, L., 1962, Physics of Fully Ionized Gases
             (New York: J. Wiley)
\bibitem[]{} St.-Louis, N., Dalton, M.J., Marchenko, S.V.,
Moffat, A.F.J., \& Willis, A.J., 1995, \apj, 452, L57
\bibitem[]{} Str\"{u}der, L. et al., 2001, \aap, 365, L18
\bibitem[]{} Turner, M.J.L. et al., 2001, \aap, 365, L27
\bibitem[]{} ud-Doula, A. \& Owocki, S.P., 2002, \apj, submitted
\bibitem[]{} Usov, V.V. 1992, \apj, 389, 635
\bibitem[]{} van den Heuvel, E.P.J., 1976, in
Structure and Evolution of Close Binary Systems (IAU Symp. 73),
eds. P. Eggleton, S. Milton, \& J. Whelan (Dordrecht: Reidel), 35.
\bibitem[]{} van der Hucht, K.A., 2001, New Ast. Rev., 45, 135 (vdH01)
\bibitem[]{} van der Hucht, K.A., Cassinelli, J.P., \& Williams, P.M., 
1986, \aap, 168, 111 (vdH86)
\bibitem[]{} Waldron, W. \& Cassinelli, J.P., 2001, \apj, 548, L45
\bibitem[]{} Wessolowski, U., 1996, MPE Report 263 (MPE: Garching), 75
\bibitem[]{} White, R.L., 1985, \apj, 289, 698
\bibitem[]{} Willis, A.J. 1996, \apss, 237, 145
\bibitem[]{} Willis, A.J. \& Stevens, I.R., 1996, \aap, 310, 577
\bibitem[]{} Wright, A.E. \& Barlow, M.J., 1975, \mnras, 170, 41
\end{thebibliography}
\end{document}